\documentclass[preprint2]{aastex}

\shorttitle{Non-Maxwellian Proton Distributions in Shocks} \shortauthors{Raymond \& Laming}

\begin{document}

\title{Non-Maxwellian Proton Velocity Distributions in Nonradiative Shocks}
\author{J.C. Raymond}

\affil{Harvard-Smithsonian Center for Astrophysics, 60 Garden Street, Cambridge, MA 02138}

\author{Philip A. Isenberg}

\affil{Department of Physics and Institute for the Study of Earth, Oceans and Space, 
University of New Hampshire, Durham}

\author{J.M Laming}

\affil{Naval Research Laboratory, Code 7674L, Washington DC 20375-5321}

\begin{abstract}
The Balmer line profiles of nonradiative supernova remnant shocks 
provide the means to measure the post-shock proton velocity distribution.
While most analyses assume a Maxwellian velocity distribution,
this is unlikely to be correct.  In particular, neutral atoms that pass
through the shock and become ionized downstream form a nonthermal
distribution similar to that of pickup ions in the solar wind.
We predict the H$\alpha$ line profiles from the combination of
pickup protons and the ordinary shocked protons, and
we consider the extent to which this distribution could affect the
shock parameters derived from H$\alpha$ profiles.  The 
Maxwellian assumption could lead to an underestimate of shock
speed by up to about 15\%.   The isotropization of the pickup ion
population generates wave energy, and we find that for the most favorable
parameters this energy could significantly heat the thermal particles. 
Sufficiently accurate 
profiles could constrain the strength and direction of the magnetic 
field in the shocked plasma, and we discuss the distortions from
a Gaussian profile to be expected in Tycho's supernova remnant.

\end{abstract}

\keywords{supernova remnants--shock waves--ISM: lines and bands--line: profiles--turbulence}

\section{Introduction}

Fast interstellar shock waves that encounter partially neutral gas 
are observable as filaments of pure Balmer line emission
if they are young compared to their radiative cooling times 
(Chevalier \& Raymond 1978; Raymond 1991; Ghavamian et al. 2001).
The Balmer lines are
produced in the thin layer just behind the shock where hydrogen atoms are
excited and ionized, and this layer is thin enough that Coulomb collisions
cannot bring different particle species into thermal
equilibrium.  Hence the Balmer lines can be used to probe the physical
processes in collisionless shocks.
 
The Balmer lines have two component line profiles.  The broad component arises from neutral H atoms 
created by charge transfer with post-shock protons, and its velocity 
width is comparable to the downstream proton thermal velocity.  The narrow component
comes from neutrals that have passed through the shock, but that have 
not been ionized by charge transfer.  Therefore, its
velocity width corresponds to the temperature of the pre-shock gas.  The
intensity ratio of the broad and narrow components depends on electron
and proton temperatures, $T_e$ and $T_p$, so that it can serve as a diagnostic 
for $T_e / T_p$ immediately behind the shock.  This is an important quantity for interpreting
X-ray spectra of SNRs and for understanding collisionless shocks.  In a few cases it has been
possible to measure the widths of UV lines of other elements, and therefore 
the kinetic temperatures, $T_i$, of other ions
\citep{ray95, laming96, ray03, korreck07}.  The overall result is that
the plasma behind relatively slow shocks ($\sim$ 300 $\rm km~s^{-1}$ )
is close to thermal equilibrium, while shocks faster than about
1000 $\rm km~s^{-1}$ are far from equilibrium, with $T_e / T_p$ $<$ 0.1
and $T_i / T_p \sim m_i / m_p$
\citep{rakowski, ghavamian07}.  Other important applications of Balmer line diagnostics
for collisionless shocks are estimates of shock speed, which can be combined
with proper motions to find SNR distances (Winkler, Gupta \& Long 2003), and inferences
of cosmic ray diffusion coefficients from the properties of shock precursors
\cite{smi94, hes94, lee07}.  

All of the current models used to interpret the Balmer line profiles assume that 
the post-shock proton velocity distribution
is Maxwellian (Chevalier et al. 1980; Lim \& Raga 1996; Laming et al. 1996),
though there 
is no solid justification for that assumption.  Coulomb
collisions are not able to bring the protons to a Maxwellian rapidly enough, and
it is not clear what sort of distribution would be produced by plasma turbulence.
Heng \& McCray (2007) have recently drawn attention
to the importance of sequential charge transfer events in determining the
profile of the broad component at shock speeds above about 2000 $\rm km~s^{-1}$,
where the charge transfer cross section changes rapidly with energy, and this
affects some of the diagnostics.  Heng et al. (2007) have extended the 
model effort to a fuller treatment of the hydrodynamics than is usually
employed, but they keep the assumption that the proton distribution is 
Maxwellian.

Ion velocity distributions directly measured in the solar wind are essentially
never Maxwellian in the vicinity of  shocks.  Ion distributions typically have
power-law tails or strong anisotropies, with beamed components upstream and
highly perpendicular enhancements downstream (Schopke et al. 1983; 1990; Gosling and Robson
1985; Thomsen 1985; Kucharek et al. 2004).
The Balmer line profiles of non-radiative
shocks provide a unique opportunity to search for non-Maxwellian velocity
distributions in astrophysical plasmas.  

In this paper we will keep the assumption that protons that pass through
the shock have a Maxwellian distribution at the temperature given by the 
Rankine-Hugoniot jump conditions (Draine and McKee 1993),
but we will add the manifestly non-Maxwellian distribution of
protons that pass through the shock as neutrals and become ionized.
We will consider the potentially observable
effects on the Balmer line profiles including line widths and centroid shifts
and how they might affect shock parameters derived from H$\alpha$ profiles.
We will also discuss the implications of magnetic field strength and direction and of plasma turbulence
on the profiles and the possibility that observed profiles could constrain the field
parameters.  We will briefly consider the implications of non-Maxwellian distributions for
the heating of electrons and minor ions.
% Excludes cosmic ray seed particle discussion.

\section{Pickup Ions}

Neutral particles that are ionized in the post-shock flow are very much like the pickup ions
(PUI) measured by spacecraft in the solar wind \cite{moebius85, gloeckler93, isenberg95}.  When neutral atoms slowly flow into the interplanetary medium,
they can be ionized by photons from the Sun, by charge transfer with solar
wind ions, or by collisions with electrons.  At that point, the new ions are streaming
with respect to the solar wind plasma at the solar wind speed, $V_{SW}$, which is much
larger than the local Alfv\'{e}n speed, $V_A$.  These ions are immediately
swept up by the magnetic field in the solar wind.  Their velocity component perpendicular to the
local magnetic field becomes a gyrovelocity around the field, which, in combination with
the instantaneous parallel component, initially forms a monoenergetic ring-beam in velocity
space.  This ring-beam is unstable, and the particles rapidly scatter toward isotropy by
interacting with ambient or self-generated waves, resulting in a velocity-space shell
\cite{sagdeev86, leeip, isenberg, bogdan}. 

In the solar wind, pickup protons are distinguished by their unusual velocity distributions,
but heavier pickup ions can also be recognized by their single
ionization states, such as $\rm He^+$ or $\rm O^+$, which stand out among solar wind
ions that are much more highly ionized.  The pickup ions add significant mass and momentum to the 
solar wind in the outer heliosphere, and the waves they generate play an important
role in heating the solar wind beyond 5 AU \cite{csmith01, isenberg03, isenberg}.  
The composition and charge state of these pickup ions indicate that they form the source 
particles for the observed anomalous component of cosmic rays (Garcia-Munoz et al. 1973, 1975; 
Fisk et al. 1974; Cummings \& Stone, 2007).  These particles must therefore be preferentially 
accelerated to several tens of MeV/nucleon at the solar wind termination shock or in the 
heliosheath beyond.  However, recent in situ observations during and after the Voyager encounters 
with the termination shock (Stone et al. 2005; Stone 2007) have shown that the energization process 
is still not well understood.  

\section{Consequences for SNR shocks}

Consider a planar shock in which the downstream magnetic field makes an angle $\theta$
with the shock normal.  Since the field component perpendicular to the flow is
compressed by the shock, $\theta$ is typically 60$^\circ$ to 85$^\circ$, though
of course pure parallel and pure perpendicular shocks maintain their field
directions.  For a strong shock with a compression ratio of 4, a neutral 
passing through the shock moves at $\frac{3}{4} V_S $ relative
to the post-shock protons.  Thus
when it becomes a pickup ion it acquires a gyro velocity 

\begin{equation}
V_{\bot} = \frac{3}{4} V_S ~sin (\theta)
\end{equation}

\noindent
and a velocity along the field direction of

\begin{equation}
V_{\|} = \frac{3}{4} V_S ~cos (\theta) 
\end{equation}

\noindent
relative to the post-shock plasma.  These monoenergetic particles  form an
unstable ring distribution in velocity space.  They can emit plasma waves
and interact with these waves to scatter into a more isotropic distribution.  
Generally, the dominant isotropization process is pitch-angle scattering through
the cyclotron resonant interaction with parallel-propagating  ion-cyclotron and
fast-mode waves (Wu \& Davidson 1972; Winske et al. 1985; Lee \& Ip 1987; see
also Zank 1999; Szeg\"{o} et al. 2000).

	The ring-beam distribution may also be subject to other plasma instabilities,
depending on the relative density and downstream conditions.  In principle, a 
downstream magnetic field nearly parallel to the flow can result in bump-on-tail 
(Gary 1978) or firehose-like instabilities (Winske et al. 1985; Sagdeev et al. 1986).
The saturation of the Landau bump-on-tail instability leaves a highly anisotropic beam 
which still scatters in pitch angle through the cyclotron resonance.  The firehose 
instability could disrupt the beam, but requires both a high density of pickup ions 
relative to the background plasma and an ionization time-scale much shorter than the 
time-scale for cyclotron resonant pitch-angle scattering.  If the downstream magnetic 
field is nearly perpendicular to the flow, a mirror-mode instability can be excited 
(Winske \& Quest 1988; McKean et al. 1995), but this instability saturates at a much 
lower level than the resonant ion-cyclotron instability (Yoon 1992), and so may be 
neglected.  In this paper, we will take the ring-beam of newly-ionized protons to 
quickly stabilize through cyclotron-resonant pitch-angle scattering.  In particular, 
we will assume the rapid formation of a bispherical distribution.

\subsection{Bispherical Distribution}

Under most conditions, a given energetic proton is cyclotron-resonant with two
parallel-propagating electromagnetic modes.  If the proton parallel speed is much
faster than the Alfv\'{e}n speed, $V_A$, both of these waves will be Alfv\'{e}n
waves -- one propagating along the field in the same direction as
the proton and the other in the opposite direction.  Resonant scattering away
from the ring-beam will result in the amplification of one of these modes and the
damping of the other.  Which mode is unstable depends on the position of the ring-beam
in velocity space, as determined by the angle of the local magnetic field to the plasma flow
direction.  The resonant interaction with either wave yields a diffusion which conserves the
proton energy in the frame of the wave phase speed, scattering the particles along
a sphere in velocity space centered on one of the points $v_{\|} = \pm V_A$, as
shown in Figure 1.  A useful
analytical result is obtained in the case where the damped
mode can be neglected and the scattering at each point in velocity space is only due to 
interactions with the unstable mode.  In this case, a steady ring-beam will be scattered 
to a bispherical distribution: a uniformly populated shell formed by the two spherical
caps which meet at the position of the original ring-beam (Galeev \& Sagdeev 1988; Williams
\& Zank 1994).

Many of the basic properties of this distribution may be obtained geometrically.  If the 
ring-beam of the newly ionized protons is located at ($V_{\|},~V_{\bot}$) as given by
equations (1) - (2), the radii of the two spherical caps are $v_{\pm}^2 = V_{\bot}^2 + (V_{\|} \pm V_A)^2$.
The area of each cap in velocity space is $a_\pm = 2 \pi v_\pm (v_\pm \mp V_{\|} - V_A)$.
Since the particles are distributed uniformly over these areas, the net
streaming speed of the bispherical shell is

\begin{equation}
v_{bulk} = \frac{1}{a_T} [ V_A (a_+-a_-) + \pi V_{\bot}^2 (v_- - v_+) ]
\end{equation}

\noindent
where the total shell area $a_T = a_+ + a_-$.  Clearly, the case of flow perpendicular
to the magnetic field, $V_{\|} = 0$, gives $v_+ = v_-$ and $v_{bulk} = 0$.  Similarly,
for parallel flow faster than the Alfv\'{e}n speed, the distribution reduces to a
single sphere of radius $V_{\|} - V_A$, and the bulk speed is slowed to $v_{bulk} = V_A$.
In general, the streaming speed of the bispherical distribution is bounded by $\pm V_A$.
Figure 2 shows this streaming speed as a function of magnetic field angle $\theta$ for
several values of the downstream field strength, taking a shock speed of 2000 $\rm km~s^{-1}$
and an upstream proton density of 1 $\rm cm^{-3}$.

These simple properties may be modified for realistic conditions.  For instance, 
dispersion of the resonant waves will systematically shift their phase speed, and so 
distort the shape of the shell away from a sphere (Isenberg \& Lee 1996).  This 
distortion can be significant if the speed difference between the neutrals and the 
downstream plasma is comparable to $V_A$.
In addition, an efficient turbulent cascade could maintain the stable wave mode intensity
despite the damping by the pickup protons.  In this case, the multiple wave-particle
interactions with both stable and unstable waves can yield a much different distribution,
and even result in particle acceleration through the second-order Fermi mechanism
(Isenberg et al. 2003; Isenberg 2005).  However, the bispherical expressions provide a
reasonable first approximation to the pickup proton distribution expected downstream of a
strong supernova shock.  In this initial study, we will retain the bispherical assumptions,
and address these simplifications in the discussion section.

\subsection{Total Proton Distribution}

At any point in the downstream plasma the velocity distribution is
the sum of the distributions of the shocked protons and the protons
formed by ionization or charge transfer in the downstream gas.  If the
preshock neutral fraction is small, the distribution is dominated
by the shocked protons, and it will be difficult to detect the effects
of the pickup protons.  These effects will be much easier to see in the shocks of
Tycho's SNR, where the neutral fraction is around 0.85 \cite{ghavamian00}
than in SN1006, where it is around 0.1 \cite{ghavamian02}.  Figure 3
shows a simple model of a shock propagating at 2000 $\rm km~s^{-1}$ into
a medium with $n_H = n_p = 0.5 \rm cm^{-3}$, roughly similar to the values
expected for Tycho's SNR. The proton density just
behind the shock is the density of thermal protons, so the increase downstream
represents the addition of pickup protons.  The neutrals immediately behind
the shock make up the slow or narrow component.  Their density drops as
charge transfer converts them to pickup protons and relaces them with fast or broad 
component neutrals.  Eventually, collisional ionization removes all neutrals,
leaving a fully ionized plasma far downstream.  The rate coefficients for charge transfer
and ionization by electrons and protons were adopted from Laming et al. (1996).
Note that this plot assumes that the pickup protons move with the same bulk speed as the
background plasma.  This will be strictly true only for a perpendicular shock,
since the scattered shell of pickup ions will generally retain some streaming
motion with respect to the thermal plasma if the field has a component along
the flow.

Figure 4 shows the thermal proton and pickup ion distributions for one choice of 
the parameters.  For the modest Alfv\'{e}n speeds expected behind
SNR shocks, the PUI distribution is not far from spherical.  Thus
the total velocity distribution shows a peak with a sharp
cutoff plus high velocity wings from the thermal distribution.

The broad components of the Balmer line profiles will reflect the proton distributions, though
they are weighted by the charge transfer cross section.  Figure 5
shows the proton velocity distribution profiles in the direction parallel
to the shock front obtained by adding the background plasma distribution
to the PUI distribution.  We have chosen this
direction because strong limb brightening is required to make the H$\alpha$
emission from a non-radiative shock bright enough that a high
S/N profile can be obtained.  The projection is obtained
by multiplying the velocity along the magnetic field direction by sin$\theta$.
If the observer is not in the plane containing the pre-shock and post-shock
magnetic field, the centroid shift will be reduced.
In this paper we do not compute Balmer line profiles, since they depend on
specific shock parameters.  Such calculations will be needed for the
interpretation of observations, but for shocks below roughly 2000 $\rm km~s^{-1}$
the variation of charge transfer cross section with velocity is weak
enough that the H$\alpha$ profile should closely resemble the proton
velocity distribution (see Heng and McCray 2006).  It should be kept in
mind, however, that the broad component of H$\alpha$ is emitted from a
region of varying pickup ion fraction, with values near zero near the shock
and approaching the pre-shock neutral fraction far downstream.  Roughly speaking,
the H$\alpha$ profile will correspond to a pickup ion fraction of about
half the pre-shock neutral fraction. 

It is apparent from Figure 5 that the departure from a Maxwellian
ought to be detectable with sufficiently high S/N data.  The difficulty
is that the narrow component, whose intensity is generally dominant,
obscures the center of the broad component profile.
The usual procedure of fitting the sum of two Gaussians to the total profile
provides enough degrees of freedom to absorb modest departures from the
assumed Gaussians, especially if the far wings of the profile and the
background level are poorly defined.  For very fast shocks, the dropoff
in charge transfer cross section with velocity may suppress the high
velocity tail in any case.

As an estimate of the error that could be made by assuming a Maxwellian
proton distribution and using the resulting
broad line width to derive a shock speed, we fit the profiles in Figure
5 with single Gaussians and compared those widths to the widths of a pure Gaussian
at the temperature expected from shock speed.
We find that the Gaussian widths estimated from the Figure 5 distributions
are as much as 14\% narrower than those predicted
for a pure thermal distribution of protons, so the shock speed
could be underestimated by 14\%.  This is the extreme case, however,
and underestimates about half that large would be typical.  These underestimates
would be partly countered if the pickup process provides additional heating to the 
plasma.

\subsection{Plasma Heating}

Another possible consequence of the pickup process is plasma heating by the waves
generated in the isotropization of the initial ring-beam of newly 
ionized protons.  The energy lost by the protons in scattering from the
ring-beam to the final nearly isotropic shell is transferred to the
resonant waves.  These waves in turn may heat the plasma, either directly or through
a turbulent cascade to dissipative modes.  In the simple bispherical picture of
section 3.1, the energy available to the waves is given by

\begin{equation}
E_w = E_o - E_{BD+} - E_{BD-}
\end{equation}

\noindent
where $E_o = mn ( V_{\bot}^2 + V_{\|}^2)/2$ is the energy in the initial ring-beam
and the energy in the bispherical distribution is 

\begin{equation}
E_{BD\pm} = \frac{nm\pi v_\pm^2}{a_T} [\frac{V_{\|}}{v_\pm}
(V_{\|} V_A \pm V_A^2 \mp v_\pm^2) + (v_\pm^2 - V_A)^2]
\end{equation}

Figure 6 shows the ratio of the total bispherical energy, $E_{BD} = E_{BD+}+E_{BD-}$
to the initial energy for various combinations of the Alfv\'{e}n speed and the
downstream magnetic field angle.  The wave energy
in (4) is essentially a maximum estimate, since the bispherical distribution
has a lower energy than the distributions obtained by including dispersive
effects or the replenishment of the stable wave modes (Isenberg \& Lee 1996;
Isenberg 2005).

The form of the plasma heating which results from the pickup proton generated
waves is an active area of research in the solar wind.   A phenomenological
model which assumes that these waves feed a turbulent cascade which dissipates by 
heating the thermal
ions has been shown to reproduce the observed proton temperatures in 
the outer heliosphere reasonably
well (Smith et al.2001, 2006; Isenberg et al. 2003; Isenberg 2005).  This 
heating can be important when
the upstream neutral fraction is large, and it may therefore affect the estimates of the
shock speed from the observed Balmer line width.

\subsection{Electron Heating}

Electron heating is observed to be very inefficient in fast shocks, so
the observed electron temperatures could provide a strong constraint 
on the wave energy even if only a modest fraction of the wave energy
is transferred to the electrons.  Observations of young SNRs show that $T_e / T_i$ is
less than 0.1 in shocks faster than about 1500 km/s (Rakowski 2005).
Ghavamian et al. (2006) propose that cosmic rays diffusing ahead of a fast
shock produce lower hybrid waves which then heat the electrons to a temperature of about 0.3 keV,
and this can reproduce the observed variation of $T_e / T_i$ with shock speed.

Alternatively, if the pickup proton ring distribution generates lower hybrid waves,
they could heat electrons.  The lower hybrid heating is inefficient unless the Alfv\'{e}n speed
is large (Omelchenko et al. 1989; Cairns \& Zank 2002), but the Alfv\'{e}n speed downstream
of SNR shocks is very poorly known.  In the absence of information about $V_A$, one cannot
make quantitative predictions.  In Tycho's SNR, which has a large neutral fraction in
the pre-shock gas, the observed low electron temperature precludes efficient transfer of energy
to the electrons if $V_A > 0.1 V_S$.

\subsection{Downstream Heating of Heavy Ions}

Other elements with ionization potentials at least as large as that of hydrogen
will be partially neutral in the pre-shock gas.  They will also be ionized and picked
up except that they will be more likely to undergo electron or proton 
collisional ionization rather than charge transfer, so the process will
occur over a thicker region behind the shock.  Thus O, N and especially
Ne and He should initially form ring distributions and be picked up by the plasma.  As with the protons,
the initial width of the ring varies as sin$\theta$ and the initial speed
along the field as cos$\theta$.

Heavy ions present in the upstream plasma can also have peculiar 
downstream distributions due to their passage
through the shock. 
They are decelerated by the electric potential jump associated with the
shock, and because of their large mass to charge ratios they are decelerated
less than the protons.  Fuselier \& Schmidt (1997) find that the initial
ring velocity in this case is 

\begin{equation}
V_{\bot} = V_s (((m/q-1)+1/16)/(m/q))^{1/2} 
\end{equation}

\noindent
for a strong perpendicular shock.  Thus we expect that heavy ions passing through
the shock will have values of $V_{\bot}$ between $V_S$ and $3 V_S / 4$.
A few observations exist to test this expectation.
The line widths of C IV and He II lines in the Hopkins Ultraviolet Telescope
spectrum of SN1006 (Raymond et al. 1995) are the same as the width of H$\alpha$
within substantial uncertainties, and the O VI line observed with FUSE is
consistent with the same width (Korreck et al. 2004).  Ghavamian et al. (2002)
find that the pre-shock neutral fraction of H is about 0.1, while that of He is at least 0.7.
Since C has a lower ionization potential than H, and O has the same ionization
potential has H, these elements also have small pre-shock neutral fractions.  Thus
C and O should have larger values of $V_{\bot}$ than H, while He should be primarily
a pickup ion distribution.  Unfortunately, the 10\% to 30\% uncertainties in the
line widths preclude a definitive  comparison, but with somewhat higher quality
profiles for the UV lines one could begin to constrain the magnetic field direction.
 
\subsection{Cosmic Ray Modified Shocks}

Except for some consideration of magnetic field amplification, the discussion above 
assumes a simple magnetohydrodynamic shock.  However, both observations and theory (e.g., 
Drury \& V\"{o}lk 1981; Malkov et al. 2000; Warren et al. 2005)
indicate that a substantial fraction of the energy dissipated in the shock, as much
as 80\%, can be converted to cosmic rays.  This results in a ``modified shock" structure
with several interesting features; 1) a particle velocity distribution such as a Maxwellian with
a power law tail, 2) a smooth transition rather than a sharp shock jump,
3) a compression ratio higher than the hydrodynamic factor of 4, and, 4) a lower
proton temperature for a given shock speed, since less energy is available to
heat the gas. 

The Balmer line profiles are not expected to reveal the non-thermal tails predicted 
for strong diffusive shock acceleration of cosmic rays, since only a very small fraction
of the particles ($\sim 10^{-3}$) are accelerated. Also, the charge transfer cross
section declines rapidly at speeds above about 2000 $\rm km~s^{-1}$ (e.g., Schultz et al. 2008), 
so the faster protons are less likely to produce broad component neutrals.  Therefore,
direct detection of the power law tail will be very difficult.
   
The smooth transition could change the profile in a manner incompatible with observations, in that
the gradually increasing temperature would give a composite H$\alpha$ profile which
is the sum of profiles formed at all the temperatures in the shock transition.  It
would probably not resemble the easily separable broad and narrow component profiles 
observed.  This difficulty would be avoided if the smooth transition occurs on a length 
scale smaller than the length scale for charge transfer, since few broad component
neutrals would form in the intermediate temperature region.  The length scale for a modified
shock is $\kappa / V_s$, where $\kappa$ is the cosmic ray diffusion coefficient.  The
charge transfer length scale is $V_s / (n_p q_{CT})$, where $q_{CT}$ is the charge transfer
rate coefficient.  Since $q_{CT} \sim 3 \times 10^{-7}~\rm cm^3 s^{-1}$ in the downstream
plasma, $\kappa$ should be smaller than about $10^{23}~\rm cm^2 s^{-1}$.  Values of
$\kappa$ of this order are required to accelerate cosmic rays to high energies within
an SNR lifetime, but they are comparable to the Bohm limit, and therefore at the
low end of the range of plausible values.

A high compression ratio, say 7 rather than the usual 4, would mean that the narrow 
component neutrals  move at $6 V_S / 7$ relative to the postshock gas, rather than 
$3 V_S / 4$, so the PUI component will have larger initial parallel and perpendicular 
velocities by 14\%. On the other hand, if a large fraction of the shock energy goes 
into nonthermal particles, the thermal speed of the protons will be reduced by a
factor $(1+P_C / P_G)^{-1/2}$, where $P_C$ and $P_G$ are the cosmic ray and gas pressures.
If $P_C$ is comparable to $P_G$, the thermal part of the line width will be seriously
affected and the shock speed will be underestimated if $P_C$ is assumed to be small.  
Most of the Balmer line filaments studied to date show very weak radio 
emission (e.g., the NW filament in SN1006 and the northern filament in the Cygnus Loop;
Ghavamian et al. 2001, 2002), so $P_C / P_G$ is probably small. 
  
\section{Application to Tycho's SNR}

Tycho's supernova remnant presents a good opportunity to search for the
effects described above because of its relatively high pre-shock neutral
fraction \cite{ghavamian00}, the excellent high and low resolution
spectra of knot g \cite{smi91, ghavamian01, lee07},
and the extensive X-ray and radio studies of both the thermal and
non-thermal shocks \cite{dickeljones, dickel, hwang, warren, bamba}.   
The preshock density is approximately
1 $\rm cm^{-3}$, the pre-shock neutral fraction is approximately 0.85 and the shock
speed is approximately 2000 $\rm km~s^{-1}$ \citep{ghavamian00, ghavamian01}.
The magnetic field is likely to be amplified in strong SNR shocks 
(e.g., Lucek \& Bell 2000; Vink \& Laming 2003),
but its strength is not accurately known.  Non-thermal synchrotron
emission from nearby parts of the blast wave of Tycho's SNR, implies that the magnetic field
is on the order of 100 $\mu$G $\equiv 1 B_{100}$ \cite{warren}, yielding an Alfv\'{e}n speed
of approximately 100 $\rm km~s^{-1}$.  The field direction is not known
with certainty, though Dickel et al. (1991) show that the polarization indicates
a predominantly radial field on scales of a few arcseconds behind the shock.
The thermal pressure implied by the Rankine-Hugoniot jump conditions with
the shock speed and the pre-shock density above yields a plasma $\beta$ of 12/$B_{100}^2$.

There is a significant shift between the centroids of the broad and narrow
components of the H$\alpha$ profiles in Tycho. 
Smith et al. (1991) and Ghavamian et al. (2001) found shifts between
the broad and narrow components of H$\alpha$ of 240$\pm$60 and
132$\pm$35 $\rm km~s^{-1}$, respectively, for two slit positions in knot g.
Smith et al. interpreted the shift as an indication that the
shock normal does not lie in the plane of the sky, so that the shift
represents a small component of the post-shock plasma speed.  This
interpretation is consistent with the observation of Lee et al. (2004), who
showed that the centroid of the narrow component is shifted with
respect to the centroid of the ambient gas in that region (though there
is some uncertainty about the size of this shift;  Lee et al. 2007).  However,
the shock normal cannot be very far from the plane of the sky, since very strong
limb brightening is required to account for the observed brightness of knot g.

Alternatively, it is possible that the shift between broad and narrow component centroids
is related to the projection of $v_{bulk}$ onto the line of sight.  
The shift is limited to approximately $V_A$, so a shift of the magnitude
measured would require that the projection of the magnetic field direction
onto the direction parallel the line of sight be fairly large, and therefore $\theta$
must be near 90$^\circ$.  Within the limits of the data now available, we
cannot tell whether the shift between broad and narrow centroids is
essentially a geometrical effect, as proposed by Smith et al. (1991) or a 
result of the pickup ion bulk speed discussed above.

A second puzzle relates to the nature of turbulence downstream from the shock.
If the 100 $\mu$G field is generated by turbulent amplification
in the shock front, it will be fairly disordered.  The 
non-resonant mechanism proposed by Bell (2004) predicts that the
scale of the turbulence is smaller than the gyroradius of cosmic ray protons
(Zirakashvili et al. 2008), and generally yields a perpendicular shock.
Compression by the shock would also make the mean field direction more perpendicular
to the shock normal.  Giacalone \& Jokipii (2007) and  Zirakashvili \& Ptuskin (2008) 
study the effects of density inhomogeneities on magnetic field generated downstream of 
the shock. Both works find significant magnetic amplification, and Zirakashvili \& Ptuskin
(2008) remark that the magnetic field component parallel to the shock normal is
more enhanced.  The interaction between the
the pickup ions and the field also tends to bend the field toward the shock
normal, and the observed field in Tycho's SNR is nearly radial at
the edge of the SNR \cite{dickel}.
A turbulent field would suggest that the pickup process occurs over
a large range of $\theta$, smearing out the profile as in Figure 5d.
Detection of a non-Gaussian profile in H$\alpha$
would provide some idea of the nature of the turbulence.  This will
require very good data and careful assessment of the instrument profile
and the background level, however, and existing data do not provide useful
constraints.

\section{Discussion}

\subsection{Caveats}

There are several qualifications to the analysis presented here. 
One is the use of PUI analysis based on Alfv\'{e}n
waves, which is appropriate for a cold plasma.  As mentioned above,
$\beta$ is around 12 for Tycho's SNR, and that will be typical for
the strong shocks seen as Balmer line filaments.  Thus, other wave modes may
be important.  It is unknown whether they will tend to 
change the directions, rather than the energies, of the protons
in the way that Alfv\'{e}n waves do.  It is also possible that they
will provide better coupling to the electrons.

Another question is whether the amplified B field behind the shock is
strongly turbulent on small scales.  If so, PUI would be generated
over a broad range of angles \cite{isenbergsw9, nemeth}, tending to wash out any
line-shift signature in
the H$\alpha$ profile.  The polarization measurements of Dickel et al
(1991) indicate that the field is reasonably well ordered on the scale
of their resolution, but it could be highly random on the 0.1$^\prime$$^\prime$
scale over which the H$\alpha$ is produced.

Finally, there is the question of momentum conservation when a significant
fraction of the downstream plasma has been picked up and streams along an oblique
magnetic field.  In this case, the thermal plasma would presumably act to 
cancel the transverse momentum, resulting in a rotation of the field
toward the shock normal.  We plan to quantitatively investigate this 
interaction in the near future.  The resolution may lie in the density gradient
of the pickup ions, but further calculations are needed.

\subsection{Implications for Balmer line filament analysis}

If the pickup ions provide a significant contribution to the H$\alpha$
profile, values of $V_s$ derived from Gaussian fits are somewhat in error.
This error would propagate into distances derived by combining shock speeds
derived from the Balmer line profiles with proper motions (e.g., Winkler,
Gupta \& Long 2003).  The modifications are probably not severe, and in cases
such as SN1006, where the pre-shock neutral fraction is only 10\% and the
contribution of pickup ions to the Balmer line profiles is only 5\%, they
would be completely negligible.  In cases where the pre-shock neutral fraction
is of order 50\%, as much as 25\% of the broad component emission could 
arise from atoms produced by charge transfer from pickup ions.  In such cases
$V_s$ would probably be underestimated.  On the other hand, waves emitted by the
pickup ions as they isotropize could heat the protons and lead to a compensating effect.

%The pickup ions might be efficiently injected into the diffusive shock acceleration
%process, providing seed particles just as heliospheric pickup ions are seed particles
%for the anomalous cosmic rays.  However, this would have implications for cosmic
%ray abundances that seem at odds with observations.   The waves produced as the
%pickup ions isotropize might heat the background plasma

If non-Maxwellian distributions can be observed by way of distortions of the H$\alpha$
profiles of non-radiative shocks, they could contain unique information about
the strength and direction of the magnetic field and the level of turbulence in
the region where the H$\alpha$ emission arises.  The most promising SNR where
non-Maxwellian distributions might be found is probably Tycho, thanks to its
large neutral fraction and relatively bright H$\alpha$ emission.

\bigskip
The authors thank Marty Lee for important suggestions.
We would also like to acknowledge very useful discussions at the Lorentz Center workshop
''From Massive Stars to Supernova Remnants" and HST Guest Observer Grant GO-10577 
and FUSE Guest Observer grant NNG05GD94G to the Smithsonian Astrophysical Observatory.  
This work was supported in part by NSF Grant ATM0635863 and NASA Grant NNX07AH75G.

\clearpage

%\begin{figure}
%\plotone{stability.eps}
%\end{figure}
 
%\figcaption{Regions in the $\theta$, $V_A$ plane that are stable and unstable
%against ion-sound production by the double-peaked velocity distribution of the
%pickup protons.
%\label{stability}}

\begin{figure}
\plotone{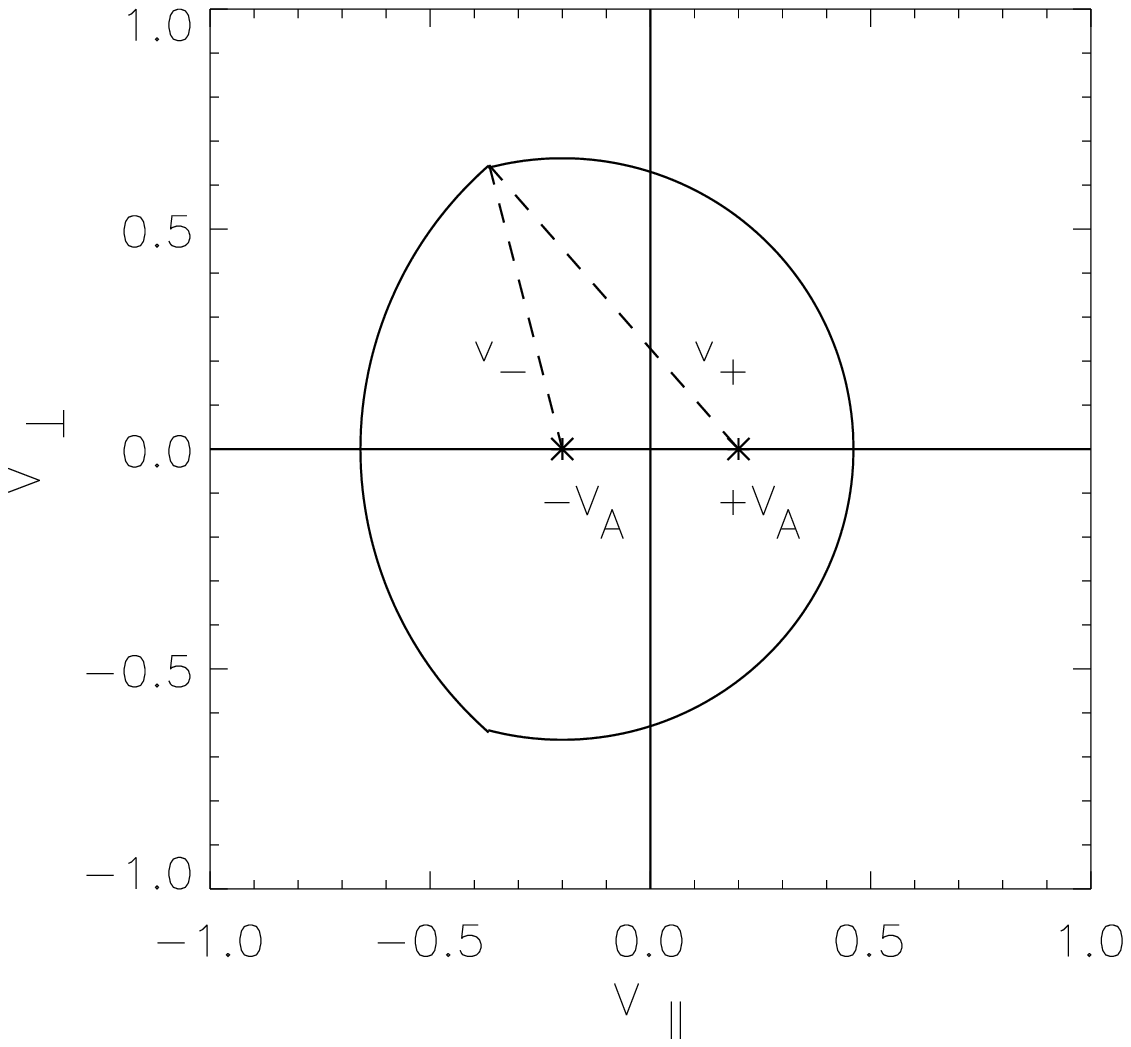}
\end{figure}
\figcaption{Bispherical distribution in velocity space where the shock speed equals 1.0.
The Alfv\'{e}n speed for this example was 0.2 $V_S$ and $\theta$ was 60$^\circ$.
\label{schem}}

%\begin{figure}
%\plotone{bisph_alf20.25.2.eps}
%\end{figure}
%\figcaption{Proton velocity distribution for an angle $\theta$=20$^\circ$
%between the field an the shock front, and Alfv\'{e}n speed $V_A$=100 $\rm km~s^{-1}$
%and a pickup ion density of 0.25 the total density.
%\label{bisphere1 }}

\begin{figure}
\plotone{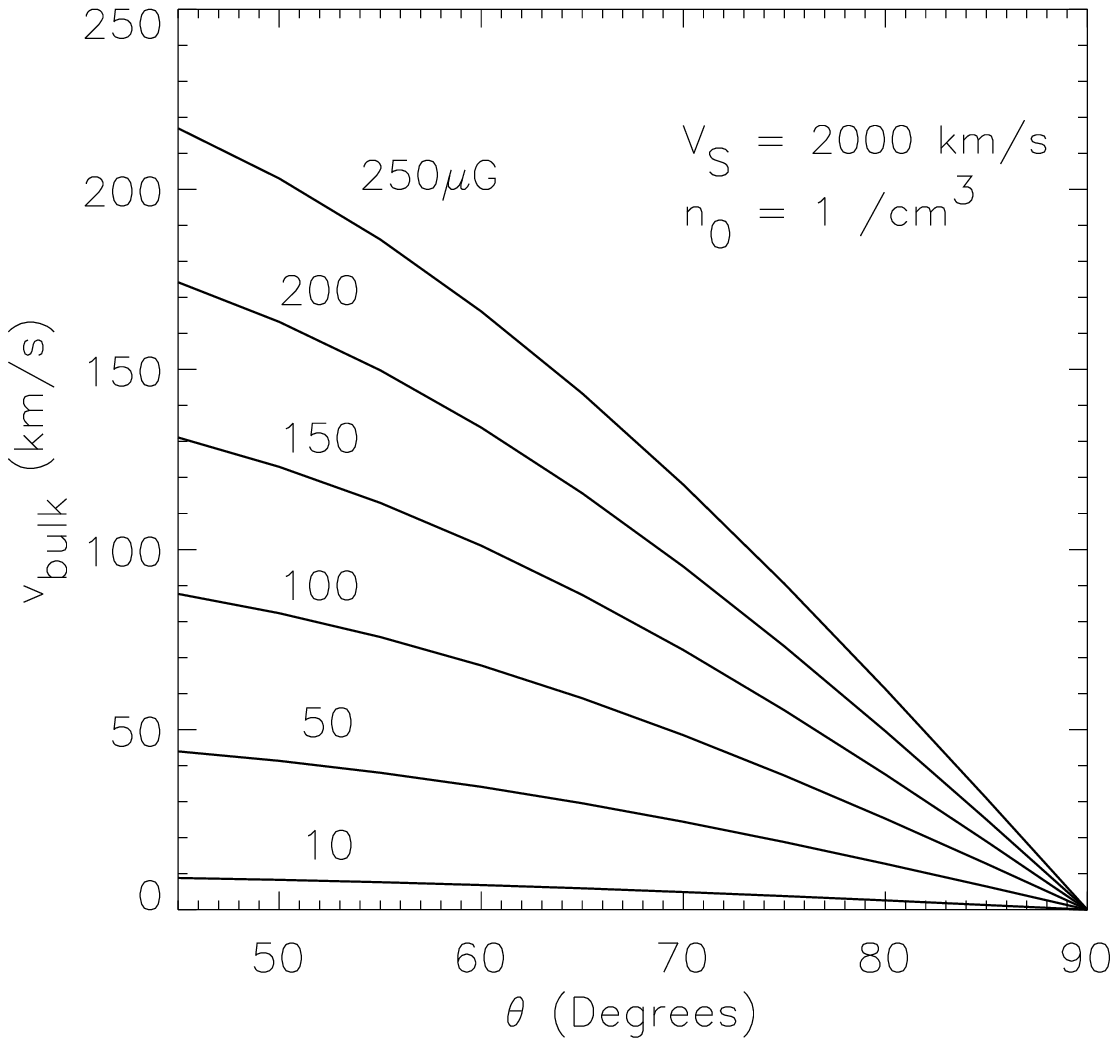}
\end{figure}
\figcaption{ Bulk velocity of a bispherical distribution along the field
direction for a 2000 $\rm km~s^{-1}$ shock with a pre-shock density of 
1 $\rm cm^{-3}$ and a post-shock density of
4 $\rm cm^{-3}$ for a range of post-shock magnetic field strengths and
angles between the shock normal and the field. The corresponding Alfv\'{e}n
speeds are 880, 710, 530, 350, 180 and 35 $\rm km~s^{-1}$.
\label{vbulk}}

\begin{figure}
\plotone{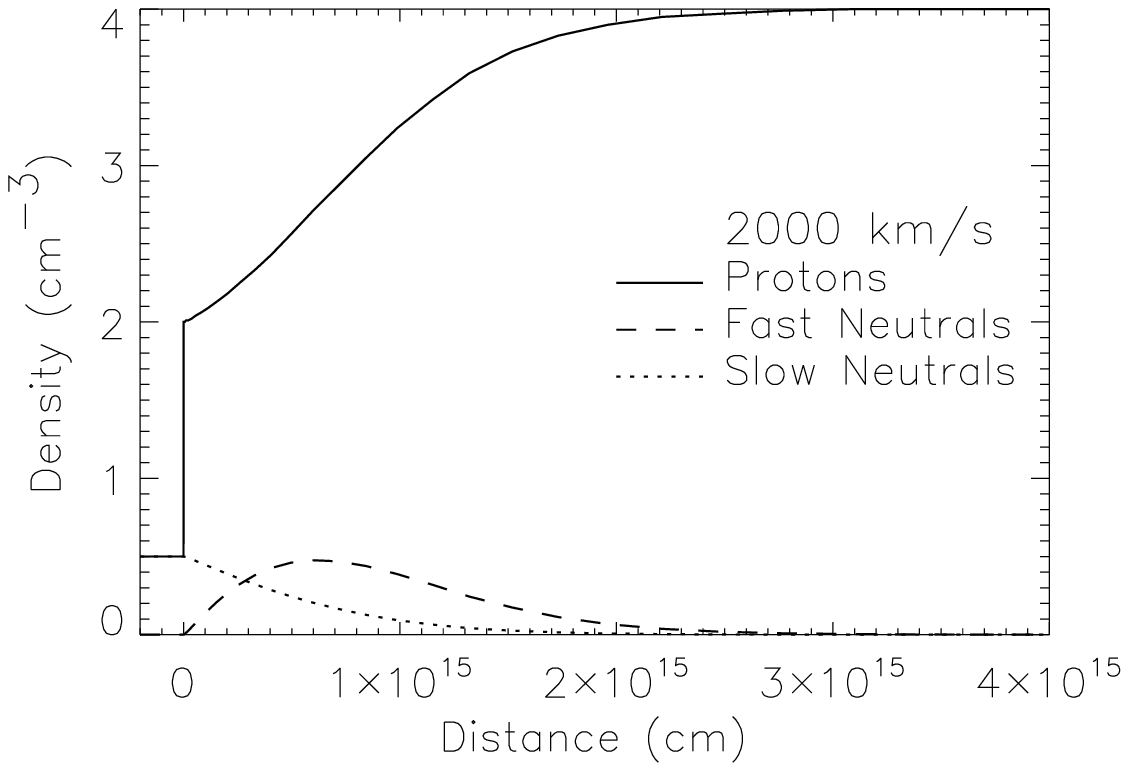}
\end{figure}
\figcaption{Variation of proton and neutral hydrogen densities behind a 2000 $\rm km~s^{-1}$
shock.  This model does not include the effects of weighting with the charge transfer
cross section or of drift on the pickup ions along the magnetic field, both of which tend
to increase the velocity of the pickup ions relative to the shock front and reduce their
density.
\label{quilha}}

\begin{figure}
\plotone{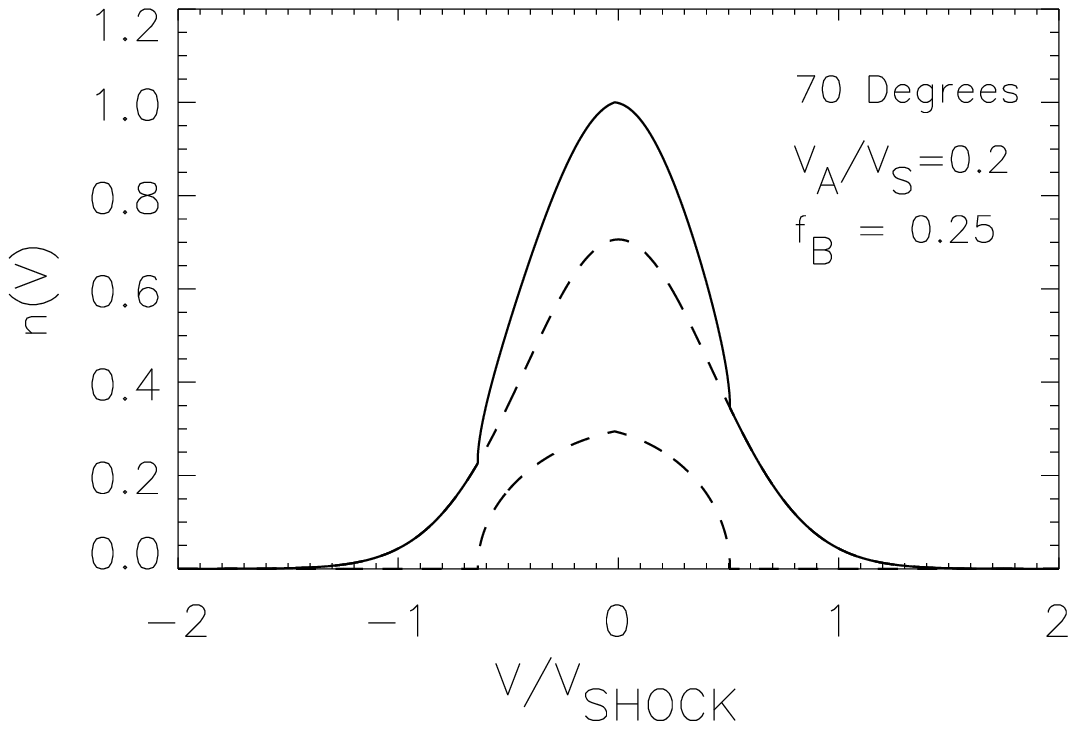}
\end{figure}
\figcaption{Proton velocity distribution for an angle $\theta$=70$^\circ$
between the field and the shock normal, an Alfv\'{e}n speed $V_A$=100 $\rm km~s^{-1}$,
and a pickup ion density of 0.25 the total density. The lower dashed line is
the bispherical distribution, the upper dashed line is the thermal proton
distribution, and the solid line is the total.
\label{bisphere1 }}

\begin{figure}
\plotone{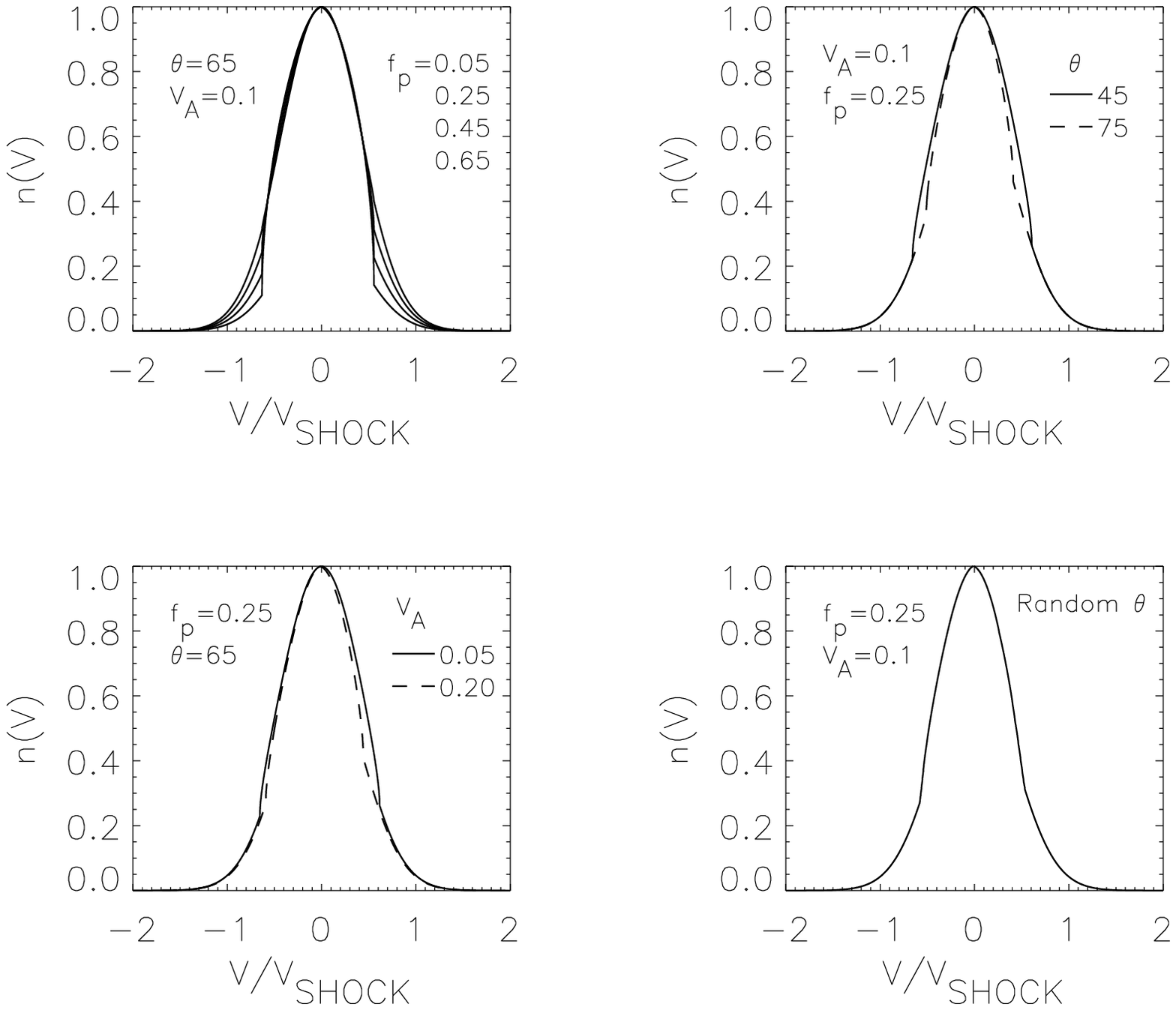}
\end{figure}
\figcaption{ Velocity distributions for various combinations of parameters.  a) Ratios of pickup ions
to thermal protons ranging from 0.05 (outermost curve)to 0.65, b) Angles between the field and the shock
normal of 45$^\circ$ (outermost curve) and 75$^\circ$, c) Alfv\'{e}n speeds of
0.05 $V_S$ (outermost curve) and 0.20 $V_S$, and d) the velocity distribution for a distribution of
angles assuming isotropic turbulence upstream and compression of $B_\bot$ by a factor of 4.
\label{params }}

\begin{figure}
\plotone{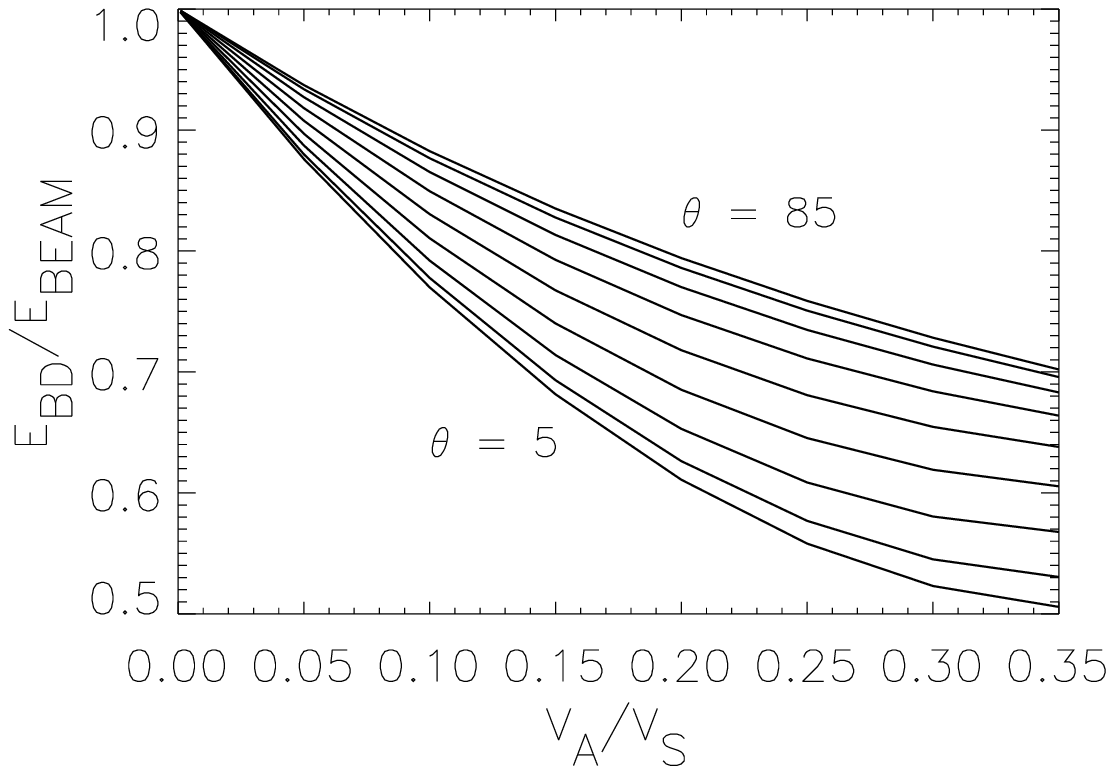}
\end{figure}
\figcaption{ Energy of the bispherical distribution as a fraction of the initial
energy of the neutral atoms.  The energy is computed in the rest frame of the post-shock
gas.  The curves correspond to values of $\theta$ of 85$^\circ$, 75$^\circ$, 65$^\circ$, 55$^\circ$,
45$^\circ$, 35$^\circ$, 25$^\circ$,  15$^\circ$ and 5$^\circ$ from top to bottom.  
\label{energy }}

\end{document}